# IS CQT MORE SUITABLE FOR MONAURAL SPEECH SEPARATION THAN STFT? AN EMPIRICAL STUDY

*Ziqiang Shi[1], Huibin Lin[1], Liu Liu[1], Rujie Liu[1], Jiqing Han[2]*

[1]Fujitsu Research and Development Center, Beijing, China
[2]Harbin Institute of Technology, Harbin, China

## ABSTRACT

Short-time Fourier transform (STFT) is used as the front end of many popular successful monaural speech separation methods, such as deep clustering (DPCL), permutation invariant training (PIT) and their various variants. Since the frequency component of STFT is linear, while the frequency distribution of human auditory system is nonlinear. In this work we propose and give an empirical study to use an alternative front end called constant Q transform (CQT) instead of STFT to achieve a better simulation of the frequency resolving power of the human auditory system. The upper bound in signal-to-distortion (SDR) of ideal speech separation based on CQT's ideal ration mask (IRM) is higher than that based on STFT. In the same experimental setting on WSJ0-2mix corpus, we examined the performance of CQT under different backends, including the original DPCL, utterance level PIT, and some of their variants. It is found that all CQT-based methods are better than STFT-based methods, and achieved on average 0.4dB better performance than STFT based method in SDR improvements.

## 1. INTRODUCTION

Multi-talker monaural speech separation has a vast range of applications. For example, a home environment or a conference environment in which many people talk, the human auditory system can easily track and follow a target speaker's voice from the multi-talker's mixed voice. In this case, if automatic speech recognition and speaker recognition are to be performed, a clean speech signal of the target speaker needs to be separated from the mixed speech to complete the subsequent recognition work. Thus it is a problem that must be solved in order to achieve satisfactory performance in speech or speaker recognition tasks. There are two difficulties in this problem, the first is that since we don't have any priori information of the user, a truly practical system must be speaker-independent. The second difficulty is that there is no way to use the beamforming algorithm for a single microphone signal. Many traditional methods, such as computational auditory scene analysis (CASA) [1, 2, 3], Non-negative matrix factorization (NMF) [4, 5], and probabilistic models [6, 7], do not solve these two difficulties well.

More recently, a large number of techniques based on deep learning is proposed for this task. These methods can be briefly grouped into three categories. The first category is based on deep clustering (DPCL) [8, 9], which maps the time-frequency (TF) points of the spectrogram into the embedding vectors, then these embedding vectors are clustered into several classes corresponding to different speakers, and finally these clusters are used as masks to inversely transform the spectrogram to the separated clean voices; the second is the (utterance level) permutation invariant training (PIT, uPIT) [10, 11], which solves the label permutation problem by minimizing the lowest error output among all possible permutations for N mixing sources assignment; the third category is end-to-end speech separation in time-domain [12, 13], which is a natural way to overcome the obstacles of the upper bound source-to-distortion ratio improvement (SDRi) in short-time Fourier transform (STFT) mask estimation based methods and real-time processing requirements in actual use.

This paper is based on the DPCL and uPIT methods [8, 9, 11], which have achieved better results than the traditional method. However, DPCL, uPIT and their most following work use STFT as front-end. Specifically, the mixed speech signal is first transformed from one-dimensional signal in time domain to two-dimensional spectrum signal in TF domain, and then the mixed spectrum is separated to result in spectrums corresponding to different source speeches by a deep clustering method, and finally the cleaned source speech signal can be restored by an inverse STFT on each spectrum. Since the distribution of the frequency components in STFT are linear, while the human auditory system is nonlinear to frequency perception, thus we hope to replace the STFT front-end with certain coefficients that can imitate human auditory system. There are two popular candidates, which are the Mel-frequency cepstral coefficients (MFCC) and constant Q transform (CQT) [14]. However, the MFCC coefficients are not suitable to be a front-end for DPCL and uPIT for two reasons, one is that it is difficult to do the inverse transform of MFCC coefficients, the other is that the sampling in the frequency-domain of MFCC is sparse. On the other side CQT with the dense coefficients are an easily reversible nonlinear transform which also very similar to the human auditory system [14]. In this work, we showed that DPCL, uPIT and their variants with CQT as front-end can achieve on average 0.4 dB better performance in separation than that with STFT as front-end.

The remainder of this paper is organized as follows: Section 2 briefly reviews the DPCL and uPIT framework. Section 3 describes the definition and implementation of CQT. The detail experimental results and comparisons are presented in Section 4 and the whole work is summarized in Section 5.

## 2. SPEECH SEPARATION WITH STFT MASK

The goal of monaural speech separation is to estimate the individual target signals in a linearly mixed single-microphone signal, in which the target signals overlap in the TF domain. The main principle of DPCL and uPIT is to use a powerful network such as LSTM to predict the mask that shows which speaker each TF bin belongs to, although their pipelines are completely different. The DPCL is to learn a high-dimensional embedding for each TF unit such that the embedding vectors belonging to the same speaker are close to each other in the embedding space, and farther otherwise [8, 9]. Then these embedding vectors will be clustered into different classes which corresponding to different speakers. While uPIT determines the best output mask automatically and then minimizes the error given the mask, which is implemented inside the network. It solves the label permutation problem and integrates speaker tracing in PIT. Thus separation and tracing can be trained in one step.

Traditional DPCL and uPIT uses STFT as front-end, however in fact STFT has linear distribution in frequency components, while CQT ensures a constant Q factor across the entire spectrum and thus gives a higher frequency resolution for low frequencies and a higher temporal resolution for high frequencies. Thus in this work we will use CQT as front-end instead of STFT in DPCL and uPIT to achieve better performance. We summarize the framework of DPCL with CQT as in Fig. 1 (and the pipeline of the uPIT method is similar, and thus is omitted here). The description of CQT will be reviewed in the next section.

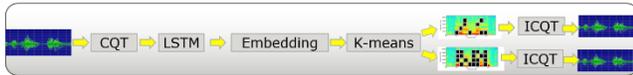

Fig. 1. The framework of DPCL with CQT

## 3. CONSTANT Q TRANSFORM

CQT was proposed by Brown [14] in 1991 to simulate the human auditory system by using a transform with fixed quality factor Q. The quality factor is a concept borrowed from the filter theory, and it is defined as the ratio of the center frequency of the filter to the bandwidth, where the bandwidth refers to the frequency at which is 3 dB less than the highest point of the filter's amplitude-frequency on the characteristic curve. For an ambiguous analogy of the transform domain, the center frequency can be considered as the frequency components of the transform domain, and the bandwidth can be considered as the frequency band of that frequency component. A series of experiments show that CQT achieves better results than STFT in music and speech analysis [14, 16, 17, 21, 22].

The original purpose of introducing CQT is to better analyze the fundamental frequency and the harmonic formant frequency position of the instrument, so as to be able to separate the sound of the musical instrument or achieve a musical instrument effect with better sound characteristics. Therefore, the bandwidth of the CQT frequency component is equivalent to a 1/24th-oct bank filter, as shown in equation (1), where $f_k$ denotes the frequency of the k-th frequency component, B denotes the 1/B octave, and $f_{min}$ denotes the minimum frequency of the CQT. The reason why B defaults always to 24 is that studies have shown that the 1/24 octave is similar to the human auditory system, but indeed for the best B is different for different applications.

$$f_k = \left(2^{1/B}\right)^k \cdot f_{min} \qquad (1)$$

Table 1. The comparison between CQT and STFT

|  | CQT | STFT |
|---|---|---|
| Frequency | $\left(2^{1/B}\right)^k \cdot f_{min}$ exponential in k | $k\Delta f$ linear in k |
| Window | Variable=$N[k] = SR \cdot Q / f_k$ | Constant = N |
| Resolution $\Delta f$ | Variable = $f_k / Q$ | Constant = SR* / N |
| $f_k/\Delta f_k$ | Constant = Q | Variable = k |
| Cycles in Window | Constant = Q | Variable = k |

The time window length corresponding to each frequency component of the STTF is same and fixed, so the frequency components are linearly distributed. Intuitively, it is only necessary to implement CQT by the STTF according to the frequency components of the CQT and take corresponding time windows, as in equation (2). The simple comparison table of CQT and STFF is shown in Table 1 [14], where SR stands for sampling rate:

$$X[k] = \frac{1}{N[k]} \sum_{n=0}^{N[k]-1} W[k,n]x[n]\exp\{-j2\pi Qn/N[k]\} \qquad (2)$$

However, the CQT directly implemented by Equation (2) cannot implement inverse transformation, which greatly limits its scope of use. Velasco et al. [16] and Holighaus et al. [17] extract the invertible CQT based on the nonstationary Gabor transform (NSGT), and combine STFT and inverse STFT to simplify the computation to improve the transform efficiency. Because of the need to implement an inverse transform, the definition of the DC component and the Nyquist CQT frequency component is increased.

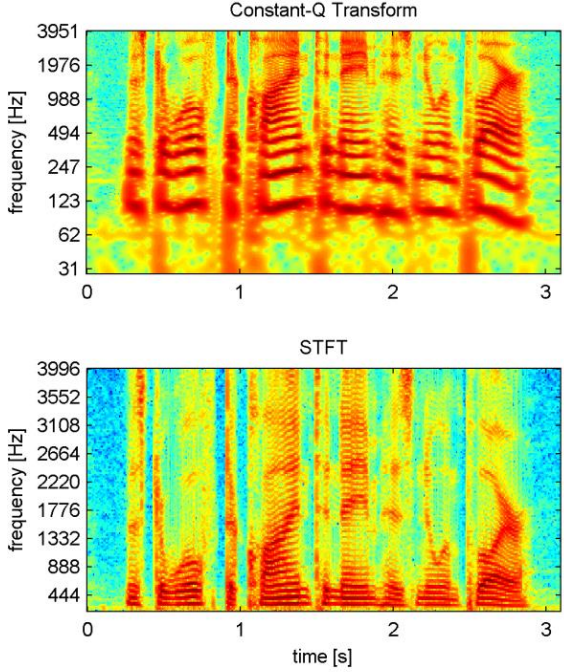

Fig. 2. Spectrograms of '22gc0103_1.9955_050c010t_-1.9955.wav' in wsj0-2mix dataset. Spectrograms computed with the STFT (top), and with the CQT (bottom).

The relationship between the center frequency and bandwidth of the CQT frequency component are shown in Table 2 [17], where ξ denotes the frequency and Ω denotes the bandwidth. $\xi_{min}$ indicates the lowest frequency of CQT, $\xi_{max}$ indicates the highest frequency of CQT, and $\xi_s$ indicates the sampling rate. Following the tradition, k denotes the index of the CQT frequency component, k = 1, ..., K, K is an integer representing $\xi_{max}$ x $\leqslant \xi_k \leqslant \xi_s / 2$, where $\xi_s / 2$ represents the Nyquist frequency. Using this method can achieve the CQT and ICQT of entire audio or partial audio. In particular, the spectrum in this section refers specifically to the frequency spectrum from the STFT where the frequency component is linear.

Table 2. The relationship between CQT center frequency and bandwidth

| k | $\xi_k$ | $\Omega_k$ |
|---|---|---|
| 0 | 0 | $2\xi_{min}$ |
| 1,.....,K | $\xi_{min} 2^{\frac{k-1}{B}}$ | $\xi_k/Q$ |
| K+1 | $\xi_s/2$ | $\xi_s - 2\xi_K$ |
| K+2,.....,2K+1 | $\xi_s - \xi_{2K+2-k}$ | $\xi_{2K+2-k}/Q$ |

According to the basic definition of CQT, the lower the frequency, the larger the bandwidth. However, for the human auditory system, only when the frequency is higher than 500Hz, it is similar to CQT, and the bandwidth below 500Hz is close to smooth. Therefore, in calculating the bandwidth of CQT, a new parameter γ is introduced, and the specific calculation formula is as shown in the following equations.

$$B_k = \alpha f_k + \gamma \quad (3)$$
$$\alpha = 2^{\frac{1}{b}} - 2^{-\frac{1}{b}} \quad (4)$$

where b represents the bandwidth of each octave equivalent filter.

To the best of our knowledge, there is no reliable open inverse CQT package in python. In order to easy in data manipulation, we develop our own inverse CQT package in python called PyEPRCQT 0.1 following the algorithm in [17].

## 4. EXPERIMENTAL RESULTS

### 4.1. Dataset and neural network

We evaluated our system on two-speaker speech separation problem using WSJ0-2mix dataset [8, 9], which contains 30 hours of training and 10 hours of validation data. The mixtures are generated by randomly selecting 49 male and 51 female speakers and utterances in Wall Street Journal (WSJ0) training set si_tr_s, and mixing them at various signal-to-noise ratios (SNR) uniformly between 0 dB and 5 dB . 5h of evaluation set is generated in the same way, using utterances from16 unseen speakers from si_dt_05 and si_et_05 in the WSJ0 dataset. To reduce the computational cost, the waveforms were down-sampled to 8 kHz.

We re-implemented the DPCL[8, 9], uPIT [10], and their variants, including CBLDNN-GAT [20] and Chimera++ [19] using STFT frontend as baselines. We only changed the frontend in these baselines to CQT and then compare the performances. The network structure and parameters of our re-implementations are basically consistent with the corresponding methods. In order to save space, we will not repeat all the configuration, interested readers please refer to the relevant literatures [8, 9, 10, 19, 20].

### 4.2. The selection of CQT parameters

There are five hyperparameters, including B, γ, window function, data block length, and minimum frequency need to select in CQT to guarantee the best performance. In order to select the most appropriate CQT configuration parameters, the SDRi upper bound of the separated speech is computed. For the case where the two speaker's speech is mixed into one, the SDRi upper bound of CQT is better than the STFT based under the same experimental conditions.

The main parameters used in CQT are B, γ, and the window functions. At the same time, the data block length and the minimum frequency of a single CQT processing will theoretically also have the influence on performance of voice separation. Therefore, we evaluate the influence of various factors on the separation performance by calculating the upper bound of the ideal SDRi. The literature [8, 9] defines the SDR ideal upper bound as follows: compute the ideal mask $a_i^{ibm} = \delta(|s_i| > \max_{j \neq i}|s_j|)$ from clean signals that are not mixed compared to the mixed signal, and then the speech signal of each speaker is separated by these ideal mask. The SDR calculated based on these separated signals is called the ideal SDR upper bound. SDRi represents the SDR of the separated voices minus the original SDR without separation. In our computation, the original SDR of the evaluation data is 0.15dB, which is consistent with the work [8, 9]. The SDRi ideal upper bound of STFT based is 13.5dB.

A grid search of the five hyperparamters, which are B, $\gamma$, window function, data block length, and minimum frequency is conducted to find the optimal parameters of CQT for speech separation, and they should be B=36, $\gamma$=20, window function is Hamming window, CQT minimum frequency is 27.5 Hz, no data block. However, considering the complexity of calculation, training the network requires the unity of data length, and because of the mistakes of previous array experiments, the standard experimental parameters of our CQT are B=36, $\gamma$=20, window function is cosine window, CQT minimum frequency It is 110Hz and the data block length is 1 second. The theoretical upper limit for SDRi is based on a CQT framework that is 1 dB higher than the STFT-based framework.

### 4.3. CQT vs. STFT based methods

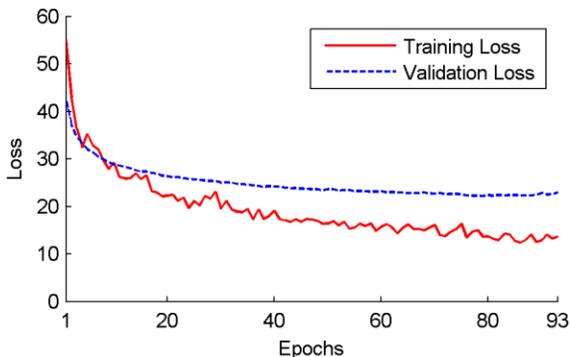

Fig. 3 Loss over epochs on the WSJ0-2mix training and validation sets with CQT based DPCL.

Based on the previous section of the experiment using CQT parameters, we trained the network and calculated the SDRi of the model. In the case of DPCL and Chimera++ the batch size is 32. Each data block contains 1 second of audio content. The CQT data block size is 126*323, which means that the frequency components of the CQT are 126 in total. Each second contains 323 CQT samples. As an example the training and validation loss in DPCL is shown in Fig. 3.

The performance is shown in the Table 3. It can be seen that the difference of the ideal SDRi upper bound between STFT and CQT based is about 1.1dB. Compared with these baselines an average increase of 0.4dB SDRi is obtained. CQT has achieved the most significant performance improvement compared with all the baseline systems.

### 5. CONCLUSIONS

In this paper we have proposed to use constant q transform (CQT) instead of short-time Fourier transform (STFT) as frontend for monaural speech separation. We give a detail description in selection of the meta-parameter of CQT in speech separation. Since CQT ensures a higher frequency resolution for low frequencies and a higher temporal resolution for high frequencies, we achieve better separation results than conventional deep clustering which uses short time Fourier transform (STFT) as front-end. It achieved on average 0.4dB better performance than STFT based method in SDR improvements.

Table 3. SDRi (dB) in a comparative study of different separation methods on the WSJ0-2mix dataset. * indicates our reimplementation of the corresponding methods.

| SDRi (dB) | STFT based | CQT based |
|---|---|---|
| DPCL [8] | 5.9 | |
| DPCL* | 10.7 | 11.1 |
| uPIT-BLSTM [11] | 10.0 | |
| uPIT-BLSTM* | 9.5 | 10.4 |
| CBLDNN-GAT [20] | 11.0 | |
| CBLDNN-GAT* | 10.8 | 11.3 |
| Chimera++ [19] | 12.0 | |
| Chimera++* | 12.2 | 12.5 |
| IBM | 13.5 | 14.6 |

### 6. ACKNOWLEGMENT

We would like to thank Jian Wu at Northwestern Polytechnical University, Yi Luo at Columbia University, and Zhong-Qiu Wang at Ohio State University for valuable discussions on WSJ0-2mix database, DPCL, and end-to-end speech separation.

# 7. REFERENCES

[1] D. Wang and G. J. Brown, Computational Auditory Scene Analysis: Principles, Algorithms, and Applications. Wiley-IEEE Press, 2006.

[2] Y. Shao and D. Wang, "Model-based sequential organization in cochannel speech," IEEE/ACM Trans. Audio, Speech, Lang. Process., vol. 14, no. 1, pp. 289–298, Jan. 2006.

[3] K. Hu and D. Wang, "An Unsupervised Approach to Cochannel Speech Separation," IEEE/ACM Trans. Audio, Speech, Lang. Process., vol. 21, no. 1, pp. 122–131, Jan. 2013.

[4] P. Smaragdis, "Convolutive Speech Bases and Their Application to Supervised Speech Separation," IEEE/ACM Trans. Audio, Speech, Lang. Process., vol. 15, no. 1, pp. 1–12, Jan. 2007.

[5] J. L. Roux, F. Weninger, and J. R. Hershey, "Sparse NMF – half-baked or well done?" Mitsubishi Electric Research Labs (MERL), Tech. Rep. TR2015-023, 2015.

[6] T. Virtanen, "Speech Recognition Using Factorial Hidden Markov Models for Separation in the Feature Space," in Proc. INTERSPEECH, 2006.

[7] M. Stark, M. Wohlmayr, and F. Pernkopf, "Source-Filter-Based Single Channel Speech Separation Using Pitch Information," IEEE/ACM Trans. Audio, Speech, Lang. Process., vol. 19, no. 2, pp. 242–255, Feb. 2011.

[8] J. R. Hershey, Z. Chen, and J. Le Roux, "Deep Clustering: Discriminative Embeddings for Segmentation and Separation," in Proc. IEEE International Conference on Acoustics, Speech and Signal Processing (ICASSP), Mar. 2016.

[9] Y. Isik, J. Le Roux, Z. Chen, S. Watanabe, and J. R. Hershey, "Single-Channel Multi-Speaker Separation using Deep Clustering," in Proc. Interspeech, Sep. 2016.

[10] M. Kolbæk, D. Yu, Z.-H. Tan, and J. Jensen, "Multitalker speech separation with utterance-level permutation invariant training of deep recurrent neural networks," IEEE/ACM Transactions on Audio, Speech, and Language Processing, vol. 25, no. 10, pp. 1901–1913, 2017.

[11] D. Yu, M. Kolbæk, Z.-H. Tan, and J. Jensen, "Permutation invariant training of deep models for speaker-independent multitalker speech separation," in Acoustics, Speech and Signal Processing (ICASSP), 2017 IEEE International Conference on, 2017, pp. 241–245.

[12] Yi Luo, and Nima Mesgarani, "TasNet: time-domain audio separation network for real-time, single-channel speech separation," in Acoustics, Speech and Signal Processing (ICASSP), 2018 IEEE International Conference on, 2018.

[13] Venkataramani, S., J. Casebeer, and P. Smaragdis. "Adaptive Front-ends for End-to-end Source Separation" , in Workshop for Audio Signal Processing, NIPS 2017.

[14] J. C. Brown, "Calculation of a constant Q spectral transform," J. Acoust. Soc. Am., vol. 89, no. 1, pp. 425–434, 1991.

[15] E. Vincent, R. Gribonval, and C. Fevotte, "Performance measurement in blind audio source separation," IEEE Trans. Audio, Speech Lang. Process., vol. 14, no. 4, pp. 1462–1469, Jul. 2006.

[16] G. A. Velasco, N. Holighaus, M. Dörfler, and T. Grill, "Constructing An Invertible Constant-Q Transform With Nonstationary Gabor Frames," Artif. Intell., pp. 93–99, 2011.

[17] N. Holighaus, M. Dörfler, G. A. Velasco, and T. Grill, "A framework for invertible, real-time constant-Q transforms," IEEE Trans. Audio, Speech Lang. Process., vol. 21, no. 4, pp. 775–785, 2013.

[18] Danwei Cai, Zhidong Ni, Wenbo Liu, Weicheng Cai, Gang Li, Ming Li, "End-to-End Deep Learning Framework for Speech Paralinguistics Detection Based on Perception Aware Spectrum." INTERSPEECH, 2017.

[19] Zhong-Qiu Wang, Jonathan Le Roux, and John R Hershey, "Alternative objective functions for deep clustering," in Proc. IEEE International Conference on Acoustics, Speech and Signal Processing (ICASSP), 2018.

[20] Chenxing Li, Lei Zhu, Shuang Xu, Peng Gao, and Bo Xu, "Cbldnn-based speaker-independent speech separation via generative adversarial training," 2018.

[21] Schörkhuber, C. and Klapuri, A. "Constant-Q transform toolbox for music processing". In Proc. 7th Sound and Music Computing Conference, Barcelona, Spain, 2010.

[22] Nagathil, A. and Martin, R. "Optimal signal reconstruction from a constant-Q spectrum". in Proc. IEEE International Conference on Acoustics, Speech and Signal Processing (ICASSP), 2012.